\documentclass[prl,twocolumn,floatfix,preprintnumbers,amsmath,amssymb,superscriptaddress]{revtex4}
\usepackage{graphicx}
\usepackage[dvips]{epsfig}
\usepackage{textcomp}
\usepackage{bm}
\newcommand\RbFe{$\rm RbFe(MoO_4)_2$}
\newcommand\KFe{$\rm KFe(MoO_4)_2$}
\newcommand\PRB[3]{Phys.~Rev.~B {\bf {#1}}, {#2} ({#3})}

\newcommand\JPSJ[3]{J. Phys. Soc. Jpn. {\bf {#1}}, {#2} ({#3})}

\begin{document}
\title{Co-existing chiral and collinear phases in a distorted triangular antiferromagnet}

\author{A.~I.~Smirnov}
\author{L.~E.~Svistov}
\author{L.~A.~Prozorova}
\affiliation{P.~L.~Kapitza Institute for Physical  Problems RAS,
119334 Moscow, Russia}

\author{ A.~Zheludev }
\author{ M.~D.~Lumsden }
\affiliation{Neutron Scattering Science Division, Oak Ridge
National Laboratory, Oak Ridge, Tennessee 37831-6393, USA}

\author{E.~Ressouche}
\affiliation{CEA-Grenoble, DRFMC-SPSMS-MDN, 17 rue des Martyrs,
38054 Grenoble Cedex 9, France.}

\author{O.~A.~Petrenko}
\affiliation{Department of Physics, University of Warwick, Coventry, CV4 7AL, UK}

\author{K.~Nishikawa}
\thanks{Present address: Hitachi, Ltd., Hitachi Systemplaza, ShinKawasaki, 890 Kashimada, Saiwai, Kawasaki,
Kanagawa, 212-8567, Japan}
\author{S.~Kimura}
\author{M.~Hagiwara}
\affiliation{Center for Quantum Science and Technology under Extreme
Conditions (KYOKUGEN), Osaka University, 1-3 Machikaneyama, Toyonaka,
Osaka 560-8531, Japan}

\author{K.~Kindo}
\affiliation{Institute for Solid State Physics (ISSP), University of
Tokyo, 5-1-5 Kashiwanoha, Kashiwa, Chiba 277-8581, Japan}

\author{A.~Ya.~Shapiro}
\author{L.~N.~Demianets} \affiliation{A.~V.~Shubnikov  Institute for
Crystallography  RAS, 117333 Moscow, Russia}
\date{\today}

\begin{abstract}
The entire magnetic phase diagram of the quasi two dimensional (2D) magnet on a distorted triangular lattice
\KFe~ is outlined by means of magnetization, specific heat, and neutron diffraction measurements. It is found
that the spin network breaks down into two almost independent magnetic subsystems. One subsystem is a collinear
antiferromagnet that shows a simple spin-flop behavior in applied fields. The other is a helimagnet that instead
goes through a series of exotic commensurate-incommensurate phase transformations. In the various phases one
observes either true 3D order or  quasi-2D order.  The experimental findings are compared to theoretical
predictions found in literature.
\end{abstract} \pacs{75.50.Ee; 76.60-k.}

\maketitle

Helical magnetic structures in frustrated spin networks have been known for several decades, see, e.g,
\cite{Yoshimori1959,Nagamiya,Zhitomirsky,Coldea}. The interest in this phenomenon was recently rekindled by a
discovery of a new class of multiferroic compounds, where ferroelectricity is inherently connected to chiral
incommensurate magnetic order (e.g. \cite{Kimura2003,Kenzelmann2005}).
 The latter enables a coupling between the
ferroelectric order parameter and a magnetic field, usually forbidden due to their respective symmetries. Perhaps
the simplest, yet fundamentally the most important model that can provide the necessary chiral state is the
two-dimensional triangular-lattice antiferromagnet (TLAF) that has a well-known ``120$^\circ$'' spin structure.

A spectacular example of a multiferroic behavior due to the TLAF
environment was recently found in the layered molybdenate \RbFe~\cite{Kenzelmann2007}. In that material triangular planes of
almost classical Fe$^{3+}$ $S=5/2$ spins  are well separated by
non-magnetic MoO$_4$ layers. The Fe$^{3+}$ subsystem is an almost
perfect realization of the TLAF model~\cite{Smirnov2007,SvistovPRB2003}. In \RbFe\ an external magnetic
field disrupts the chiral magnetic structure and thereby removes
the multiferroic effect. Even more interesting behavior can be
expected in materials with a slightly distorted TLAF spin network.
The distortion will unbalance the 120$^\circ$ structure and
produce an incommensurate planar spin spiral. In an external
magnetic field, theory predicts a multitude of exotic
incommensurate and commensurate, chiral and centrosymmetric
phases~\cite{Nagamiya}. The chiral states could, at least in
principle, possess multiferroic properties. The focus of this work
is \KFe, a material structurally similar to \RbFe\, but one that
actually realizes the {\it distorted} TLAF model. Using a
combination of experimental techniques, we find a remarkably
complex magnetic phase diagram with {\it co-existing} collinear
and helical structures, ordered in either three or two dimensions.

\begin{figure}
 \includegraphics[width=1.0\columnwidth]{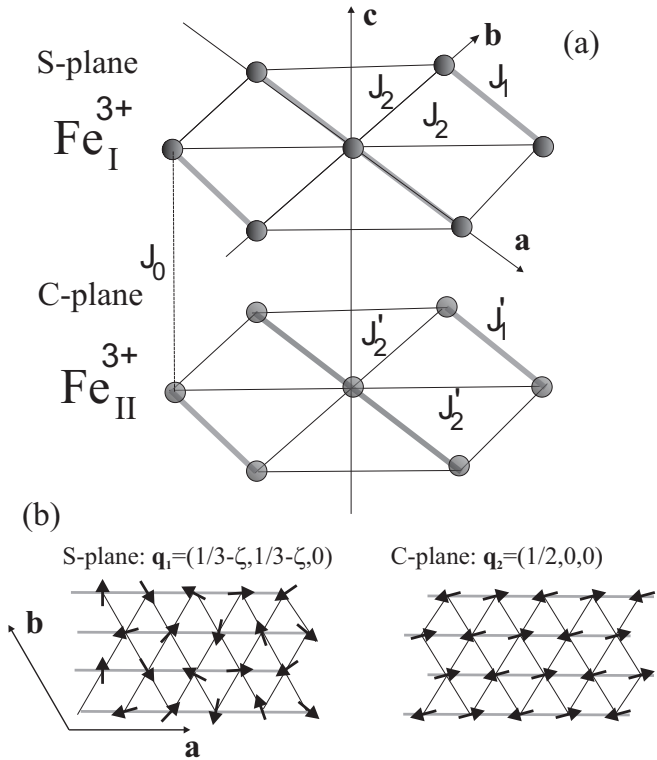}
 \caption{(a) Arrangement of magnetic ions in the crystal structure
 of KFe(MoO$_4$)$_2$ showing two inequivalent Fe$^{3+}$ planes.
 (b) Schematic representation of the zero-field magnetic
 structure.}
  \label{fig:structure}
\end{figure}

At high temperatures \KFe\  has a perfect triangular spin network due to the crystal symmetry of $D_{3d}^3$ group
with $a=5.66$~\AA\ and $c=7.12$~\AA. The distortion occurs as a result of a crystallographic phase transition
at $T=311$~K~\cite{Klevtsova,Smolenski}.  The low-temperature phase is monoclinic, with a doubling of the period
along the $c$ axis. Each Fe-plane becomes a distorted triangular lattice with two unequal exchange constants
$J_1$ and $J_2$ (Fig.~\ref{fig:structure}). Moreover, the adjacent Fe-layers become crystallographically
inequivalent, with exchange constants $J_1'$ and $J_2'$. The actual lattice distortion is too small to be
detected with the resolution of our experiments, and we shall henceforth adopt a hexagonal lattice notation.  A
previous ESR study~\cite{SvistovJETPL} of \KFe\ came to a seemingly paradoxical conclusion: at low temperatures,
in zero field,  a helical spin structure coexists with a collinear state. It was hypothesized that the two type
of magnetic order reside almost independently in the two inequivalent types of distorted triangular spin
lattices, referred to as ``S-layers'' and ``C-layers'', respectively.

\begin{figure}
\centering \epsfig{file=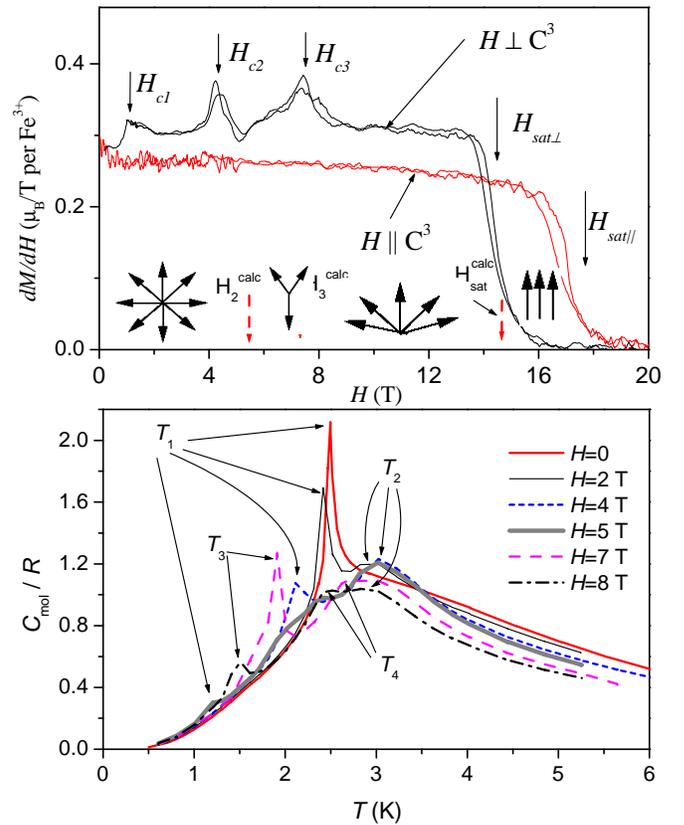, width=1.0\columnwidth, clip=}
 \caption{(Color online)(a) Field derivative of the magnetization measured at $T=1.6$~K. Solid arrows are experimentally observed anomalies.
 Dashed arrows and spin diagrams are explained in the text. (b)  Specific heat measured on a 1mg  \KFe\ single
 crystal sample in magnetic fields applied in the $(a,b)$ plane.}
  \label{fig:bulk}
\end{figure}

To verify this bold assumption we performed magnetic neutron
diffraction experiments using single-crystal samples from the same
batch \cite{SvistovJETPL}. The crystals are transparent thin
plates, of a natural triangular shape, typically 1-5~mg, with the
planes of natural growth perpendicular to the 3-fold axis. The
data were taken on the HB-1 and HB-1A 3-axis spectrometers at ORNL
operating in 2-axis mode, using a Pyrolitic Graphite PG(002)
monochromator to select $\lambda=2.46$~\AA\ for HB-1 and
$\lambda=2.37$~\AA\ for HB-1A. At low temperatures, two sets of
magnetic Bragg peaks emerge, with propagation vectors
$(1/3-\zeta,1/3-\zeta, 0)$, $\zeta=0.038$, and $(1/2,0,0)$,
respectively.  The observed ordering temperatures for the two sets
of reflections are identical within experimental accuracy:
$T_\mathrm{N}=2.4$~K. An analysis of 19 inequivalent Bragg
reflections in the $(h,k,0)$ plane at $T=1.5$~K revealed that the
$(1/3-\zeta,1/3-\zeta, 0)$ peaks can be entirely accounted for by
a planar helimagnetic state, with spins rotating in the $(a,b)$
plane. At the same time, the 14 inequivalent sets of
$(1/2,0,0)$-type Bragg intensities measured in the $(h,k,0)$ plane
are consistent with a collinear AF spin arrangement, with spins in
the $(a,b)$ plane and forming a small angle of $15^\circ$ with the
$a$ axis. Thus the diffraction data confirm the original two-layer
model depicted in Fig.~\ref{fig:structure}b. As discussed in
Ref.~\cite{SvistovJETPL}, the drastically different spin
arrangement in C- and S-layers can be accounted by the difference
in the corresponding ratios $R=J_1/J_2$ vs. $R'=J_1'/J_2'$: theory
predicts a switch from a helimagnetic to a collinear state at
$R>2$~\cite{Nagamiya}.

\begin{figure}
 \includegraphics[width=1.0\columnwidth]{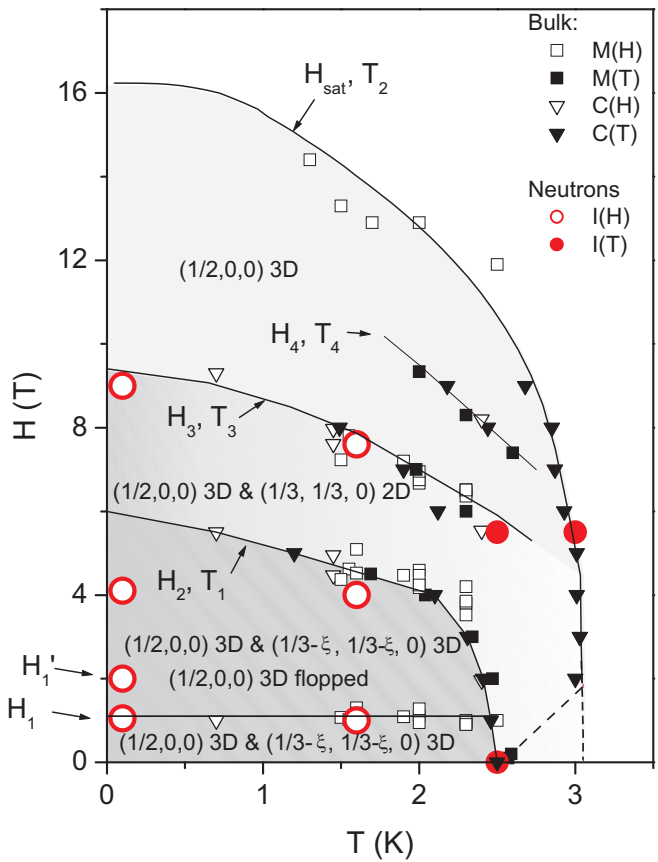}
 \caption{(Color online)Cumulative magnetic phase diagram
 of \KFe\ for a magnetic field applied in the
$(a,b)$ plane. }
  \label{fig:PhD}
\end{figure}

The complexity of the $H-T$ phase diagram of \KFe\ was revealed in
bulk magnetic and calorimetric measurements. Steady-state
magnetization data were collected in fields up to 12~T using an
Oxford Instruments vibrating sample magnetometer and specific heat
data were collected on a Quantum Design PPMS at Warwick
University. High-field magnetization data were taken in the fields
up to 25~T using a pulsed magnet at the KYOKUGEN center. Typical
experimental $C(T)$ and $dM/dH$ curves are shown in
Fig.~\ref{fig:bulk}.  In zero field we observe a sharp specific
heat anomaly at $T_1=2.5$~K. In a magnetic field applied in the
$(a,b)$ plane this anomaly shifts to lower temperatures and
survives up to $H=5$~T. For $H>2$~T an additional peak is observed
at $T_2>T_1$. Beyond $H>5$~T the $T_1$ anomaly is replaced by a
new feature at $T_3$. The latter also moves to lower
temperatures with increasing field. Yet another maximum in
specific heat is observed at $T_4<T_2$ in the high field regime.
In the magnetization data, for a field applied along the $c$ axis,
the only observed feature is saturation at
$H_{\mathrm{sat}\parallel}=16.9$~T. However, for a field in the
$(a,b)$ plane, the magnetization curves show four distinct
anomalies. A jump of the field derivative of magnetization $dM/dH$
at $H_{1}\sim 1.2$~T is followed by sharp maxima at $H_2 \simeq 4.5$~T and $H_3 \simeq 7.5$~T.
Finally, a saturation is reached at $H_{\mathrm{sat}\perp}=14.5$~T.
For different orientations of the field within the $(a,b)$ plane, and for samples with varying
populations of the three crystallographic domain types, the
measured magnetization curves are very similar. In particular, the
angular variation of $H_2$ and $H_3$ is 10\% and 7\%,
respectively. Some samples show a smeared peak in $dM/dH$ in the
field range above 8~T in the temperature interval 2 K$<T<2.5$~K.
This anomaly, presented by the magnetic field $H_4$ in Fig.~\ref{fig:PhD}, was not visible in the sample used for pulse
measurements. The magnetic and thermodynamic anomalies described
above, together with the $M(T)$ data from
Ref.~\cite{SvistovJETPL}, allow us to reconstruct the entire $H-T$
phase diagram, as shown in Fig.~\ref{fig:PhD}. The low-field part
of the phase boundary at $T_2$ is uncertain because of the absence
of this anomaly at $H<2$~T.

\begin{figure}
  \includegraphics[width=1.0\columnwidth]{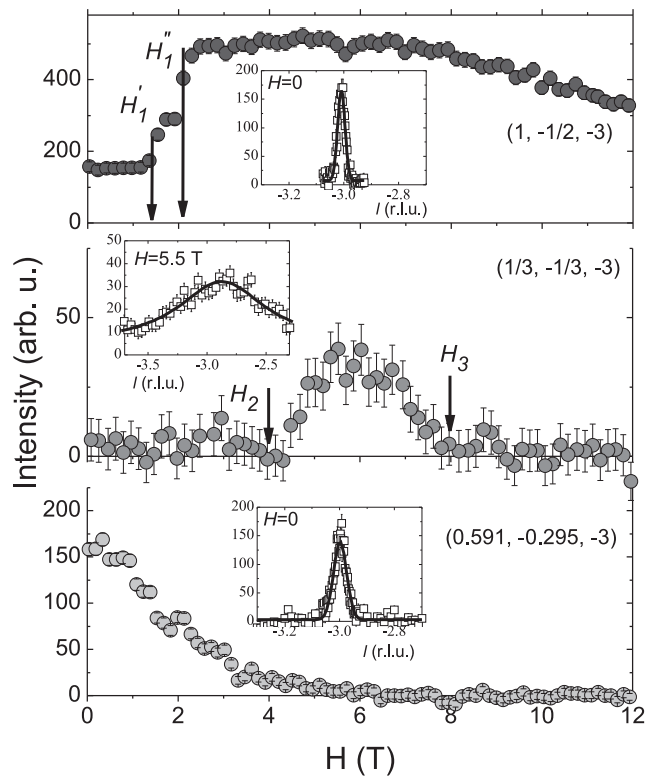}
 \caption{Main panels: field dependencies of magnetic Bragg
 intensities measured in \KFe\ at $T=100$~mK. Insets: typical
 $l$-scans measured across the corresponding reflections.}
  \label{fig:neutron}
\end{figure}

The propagation vectors in the various phases  were determined in neutron experiments on the D23 lifting counter
diffractometer at ILL using a PG(002) monochromator and $\lambda=2.38$~\AA\ neutrons. The field was applied along
the $b$ axis (in hexagonal notations). Sample environment was a dilution refrigerator. Typical measured field
dependencies of Bragg intensities are shown in Fig.~\ref{fig:neutron}. At $T=100$~mK the intensities of
commensurate $(1/2, 0, 0)$-type peaks go through two consecutive jumps at $H_1=1.3$~T and $H_1'=2.1$~T. These
transitions seem to have no effect on the incommensurate reflections. At $T=1.5$~K a single intensity jump is
detected at $H_1=1$~T. As previously discussed in \cite{SvistovJETPL}, the transition is to be associate with a
spin flop in the C-planes, all moments rotating to be perpendicular to the applied field. The additional
transition seen at low temperature at $H_1'$ requires further investigation.  At higher fields all the action
occurs within the S-planes. The intensity of the incommensurate $(1/3-\zeta,1/3-\zeta, 0)$-type reflections
decreases and vanishes beyond $H_2\sim 4$~T.  Within experimental resolution the value of the magnetic
propagation vector is field-independent. Beyond $H_2$ the $(1/3-\zeta,1/3-\zeta, 0)$-type peaks are replaced by
{\it commensurate} reflections of type $(1/3,1/3, 0)$. The latter first increases in intensity, peaks at around
6~T, and decreases at higher fields to vanish at $H_3\sim 8-9$~T. At still higher fields, no magnetic reflections
were found on either the $(h,h,0)$, $(1/3-\zeta,1/3-\zeta, l)$, or $(1/3,1/3, l)$ reciprocal-space rods. At the
temperature of neutron measurements ($T=100$~mK and $T=1.5$~K) we found no signature of a high-field transition
that could be associated with the $T_4$-anomaly described above.

A remarkable feature of the neutron data is the different dimensionality
of magnetic ordering in the different phases. Scans across the $(1/2, 0,
0)$-type and $(1/3-\zeta,1/3-\zeta, 0)$-type reflections are
resolution-limited along the $h$-, $k$- and $l$- directions. In contrast,
the $(1/3,1/3, 0)$-type peaks in the regime $H_2<H<H_3$ are actually Bragg
rods parallel to the $c$ axis, stretching across much of the Brillouin
zone in the $l$-directions. The corresponding $c$-axis correlation length
is only 6 lattice units. The measured correlation length within the
$(a,b)$ plane is much larger, about 100 lattice units. The 2D character of
the ordering persists in samples cooled in a 6~T applied field. The
transition to the short range ordered state at $T_2$ naturally
demonstrates a smeared $C(T)$ anomaly in contrast to a sharp peak at the
transition to the 3D ordered phase at $T_1$.

Guidance to understanding the complex phases realized in the $S$-planes
can be drawn from the theoretical work of Ref.~\cite{Nagamiya}. At first,
we can estimate the relevant exchange parameters from the saturation
fields and susceptibility. Because the neutron reflections observed in
high fields correspond to C-planes, we assume the saturation is associated
with this kind of planes, though S-planes, naturally, also give a
contribution to the magnetization. For the C-planes, the saturation fields
are given by: $g\mu_\mathrm{B}H_\mathrm{sat\perp} = 8(J_{1}'+J_{2}')S$ and
$g\mu_\mathrm{B}H_\mathrm{sat\parallel} = 8(J_{1}'+J_{2}')S+2DS$,
respectively, where $D$ is the single-ion easy-plane anisotropy defined as
in \cite{SvistovPRB2003}. Using the observed values of
$H_\mathrm{sat\perp}$ and $H_\mathrm{sat\parallel}$ we get $J_{1}'+J_{2}'=
0.96$ K and $D'=0.32$ K.  Beyond the spin-flop at $H_1$, the C-planes
contribution to magnetic susceptibility should be constant: $\chi_C=
g^2\mu_\mathrm{B}^2/[8(J_1'+J_2')]$. Subtracting this value from  $dM/dH$
data for $H_{c1}<H<H_{c2}$ we obtain the susceptibility of the $S$-planes:
$\chi_S\simeq 0.12~\mu_\mathrm{B}/$T per Fe$^{3+}$ ion. Using the latter
value and the measured $\zeta=0.038$, by applying the equations in
Ref.~\cite{Nagamiya}, we get $J_1=0.37$~K; $J_2=0.69$~K. As a
self-consistency check, the experimental ratio $R=0.53$ warrants a
helimagnetic ground state for the S-layers in zero field~\cite{Nagamiya}.

Now, the measured values  $\zeta$ and  $H_{sat\perp}$ can be applied to reconstruct the phase transitions in the
$S$-planes. The critical fields $H_2^\mathrm{calc}=5.7$~T, $H_3^\mathrm{calc}=7.2$~T, calculated  by use of
$\zeta$  and $H_{sat}$ following the theory ~\cite{Nagamiya}, are shown in dashed lines in Fig.~\ref{fig:bulk}a.
As indicated by the arrow diagrams, and in perfect agreement with the diffraction experiments, at low fields one
expects an incommensurate spiral structure confined to the $(a,b)$ plane. Beyond the phase transition at
$H_2^\mathrm{calc}$ the spins should form a commensurate 3-sublattice configuration with the 2D propagation
vector  $(1/3,1/3)$, as observed experimentally. At still higher fields, beyond $H_3^\mathrm{calc}$, the
incommensurate state is expected to be restored. The spins will form a modulated fan-type structure, this time
oscillating near the field direction in a small angular interval, still remaining within the $(a,b)$ plane. Note
that neutron diffraction failed to detect any incommensurate peaks beyond $H_3^\mathrm{calc}$. This implies that
either the system remains disordered,
or that the ordering vector is outside our search range in reciprocal space and was simply overlooked. Besides,
the flop of the spin plane perpendicular to the field direction is expected \cite{Nagamiya} at $H=H_{sf}=6.5$~T.
Instead of this instability we observe a  lost of the magnetic Bragg peaks from S-planes.

It is remarkable that even as the S-plane go through a series of phase transitions, the C-planes remain
unaffected, and vice versa. This implies that magnetic interactions between each type are direct, rather than
mediated by the intercalated layers of the other type.  These long-range interactions are likely to be of dipolar
origin.

In summary, we have demonstrated that \KFe\ realizes not just one, but {\it two} co-existing and decoupled
instances of the distorted TLAF model. While one has a commensurate ground state and rather simple behavior in
applied fields, the other features a chiral incommensurate structure and a series of exotic high-field phases.
Future work should consider the possibility of multiferroic behavior associated with the S-layers.

Research at ORNL was funded by the US Department of Energy, Office of Basic Energy Sciences - Materials Science,
Contract No. DE-AC05-00OR22725 with UT-Battelle, LLC. The work at Kapitza Institute is supported by the Russian
Foundation for Basic Research. The work at Warwick University is supported by EPSRC grant.
A portion of this work was done as a part of a Foreign Visiting Professor Program in KYOKUGEN, Osaka University.

\end{document}